\documentclass[aps,showpacs,preprintnumbers,amsmath, amssymb]{revtex4}

\usepackage{float}
\usepackage{graphics,epsfig}
\usepackage{graphicx}
\usepackage{dcolumn}
\usepackage{bm}

\begin{document}
\baselineskip=0.8 cm

\title{{\bf Hair formation in the background of noncommutative reflecting stars}}
\author{Yan Peng$^{1}$\footnote{yanpengphy@163.com},Qiyuan Pan$^{2}$\footnote{panqiyuan@126.com},Shuangxi Yi$^{3}$\footnote{yisx2015@qfnu.edu.cn}}
\affiliation{\\$^{1}$ School of Mathematical Sciences, Qufu Normal University, Qufu, Shandong 273165, China}
\affiliation{\\$^{2}$ Key Laboratory of Low Dimensional Quantum Structures and Quantum Control of Ministry of Education,
Synergetic Innovation Center for Quantum Effects and Applications,
and Department of Physics, Hunan Normal University, Changsha, Hunan 410081, China}
\affiliation{\\$^{3}$ School of Physics and Physical Engineering, Qufu Normal University, Qufu,Shandong 273165, China}

\vspace*{0.2cm}
\begin{abstract}
\baselineskip=0.6 cm
\begin{center}
{\bf Abstract}
\end{center}

We investigate scalar condensations around noncommutative compact reflecting stars.
We find that the neutral noncommutative reflecting star cannot support
the existence of scalar field hairs.
In the charged noncommutative reflecting star spacetime,
we provide upper bounds for star radii. Above the bound, scalar fields cannot exist
outside the star. In contrast, when the star radius is below the bound,
we show that the scalar field can condense. We also obtain the largest
radii of scalar hairy reflecting stars.

\end{abstract}

\maketitle
\newpage
\vspace*{0.2cm}

\section{Introduction}

The classical no scalar hair theorem provides insights into physical
properties of black holes \cite{Bekenstein,Chase,C. Teitelboim,Ruffini-1}.
According to this theorem, asymptotically flat black holes
cannot support static massive scalar fields, for references please refer to \cite{Hod-1}-\cite{Brihaye}
and reviews can be found in \cite{Bekenstein-1,CAR}.
It was usually believed that these no scalar hair behaviors are
due to the existence of black hole horizons,
which can inevitably absorb matter and radiation fields.

Whether the no scalar hair behavior exists in the horizonless spacetime
is a question to be answered. Recently, it was found that the
no scalar hair behavior appears in the background of asymptotically flat
horizonless neutral compact reflecting stars \cite{Hod-6}.
Moreover, it was shown that the massless scalar field nonminimal coupled
to the gravity cannot condense around the asymptotically flat neutral compact reflecting star \cite{Hod-7}.
When extending the discussion to spacetimes with a positive cosmological constant,
it was proved that massive scalar, vector and tensor hairs all die out
outside asymptotically dS neutral reflecting stars \cite{Bhattacharjee}.

However, the no hair theorem obtained in the regular neural gravity
could be challenged in the charged spacetime.
In the case of regular charged reflecting shell,
it was analytically found that the scalar hair can form outside the shell
when the shell radius is below an upper bound \cite{Hod-8,Hod-9,Yan Peng-1}.
It was also shown that regular charged reflecting stars
can support the existence of the scalar hair if the star radius is below
an upper bound \cite{Hod-10,Yan Peng-2,Yan Peng-3,Yan Peng-4,Yan Peng-5}.
We mention that there are usually radii bounds,
above which the scalar field cannot condense outside the objects
\cite{Yan Peng-1,Hod-10,Yan Peng-2,Yan Peng-3,Yan Peng-4,Yan Peng-5}.

Recently, the noncommutative gravity has attracted a lot of attentions based on the belief that
the noncommutativity will appear at the Planck scale, where usual semiclassical approaches fail.
The area law of noncommutative black holes has been
studied \cite{PN1,PN2,PN3,PN4,SG,PN5,PN6}.
Another important motivation to study noncommutative theories is
it's natural emergence in string theory and some surprising consequences \cite{EWN,NSE,SMN,JGT,DB,HDN,KM}.
As we know, the discussions on no hair theorem in regular gravities were carried out in the commutative spacetimes.
So it is interesting to extend the discussion to noncommutative backgrounds.
And it is also meaningful to construct regular hairy configurations
in the noncommutative gravity.

The rest of this work is organized as follows. In section II,
we construct the gravity system composed of a scalar field
and a charged reflecting star in the noncommutative geometry.
In part A of section III, we analytically obtain an upper bound for
the charged scalar hairy star radius.
In part B of section III, we numerically study scalar
hairy star solutions.
The last section is devoted to the conclusion.

\section{The gravity system in the noncommutative geometry}

The idea of noncommutative spacetime was firstly introduced by Snyder \cite{N1}
and such a noncommutative structure
naturally emerges in string theory \cite{N2}.
There are mainly two ways to obtain the noncommutative quantum field theory: the Weyl-Wigner-Moyal
star product approach and the coordinate coherent state approach \cite{N3}.
Most recently, inspired by the coordinate coherent state approach,
P.Nicolini and other authors obtained noncommutative black hole metrics \cite{PN5,SG,PN6},
where the noncommutativity is introduced by
writing down the Gaussian distribution of mass and charge densities
in the form
\begin{eqnarray}\label{AdSBH}
\rho_{\theta}^{(M)}=\frac{M}{(4\pi \theta)^{3/2}}exp(-\frac{r^2}{4\theta}),
\end{eqnarray}
\begin{eqnarray}\label{AdSBH}
\rho_{\theta}^{(Q)}=\frac{Q}{(4\pi \theta)^{3/2}}exp(-\frac{r^2}{4\theta}).
\end{eqnarray}
Here $\theta$ is the parameter used to describe the
noncommutativity of the spacetime.
And the mass M and charge Q diffuse throughout a region of
linear size $\sqrt{\theta}$.
When the object size is below the characteristic length $\sqrt{\theta}$,
noncommutativity becomes non-negligible.
And in the case of $\theta\rightarrow 0$,
the model returns to the commutative one.

We can search for solutions of Einstein's equations
with the mass and charge distribution (1) and (2).
The metric of the spherically symmetric noncommutative compact star is \cite{PN5,SG,PN6}
\begin{eqnarray}\label{AdSBH}
ds^{2}&=&-g(r)dt^{2}+g^{-1}dr^{2}+r^{2}(d\varphi^2+sin^{2}\varphi d\phi^{2}).
\end{eqnarray}
The solution is $g(r)=1-\frac{4M}{\sqrt{\pi}r}\gamma(\frac{3}{2},\frac{r^2}{4\theta})
+\frac{Q^2}{\pi r^2}[\gamma^{2}(\frac{1}{2},\frac{r^2}{4\theta})
-\frac{r}{\sqrt{2\theta}}\gamma(\frac{1}{2},\frac{r^2}{2\theta})
+\sqrt{\frac{2}{\theta}}r\gamma(\frac{3}{2},\frac{r^2}{4\theta})]$,
where $\gamma$ is the incomplete
gamma function in the form $\gamma(n,z)=\int_{0}^{z}t^{n-1}e^{-t}dt$.
Here M and Q are interpreted as the star mass and star charge respectively.
We also define $r_{s}$ as the regular star radius and
there is $g(r)>0$ for $r\geqslant r_{s}$.

We study scalar condensations in the noncommutative charged reflecting star background.
And the Lagrange density with scalar fields coupled to
the Maxwell field reads
\begin{eqnarray}\label{lagrange-1}
\mathcal{L}=-\frac{1}{4}F^{MN}F_{MN}-|\nabla_{\mu} \psi-q A_{\mu}\psi|^{2}-m^{2}\psi^{2}.
\end{eqnarray}
Here $\psi(r)$ is the scalar field and $A_{\mu}$ corresponds to the Maxwell field.
We also label q and m as scalar field charge and scalar field mass respectively.

We assume that the Maxwell field has only
the nonzero $t$ component in the form $A_{t}=\phi(r)dt$.
Then the equation of the Maxwell field is \cite{e1,e2,e3,e4,e5,e6}
\begin{eqnarray}\label{BHg}
\phi''+\frac{2}{r}\phi'-\frac{q^2\psi^{2}(r)\phi(r)}{2g(r)}=0.
\end{eqnarray}
In this work, we neglect the scalar field's backreaction
on the charged star. So the third term of equation (5)
disappears and the electric potential is not affected by the non-commutativity.
We take the electric potential in the form $A_{t}=-\frac{Q}{r}$ \cite{Hod-10,Yan Peng-2}
and the scalar field equation is
\begin{eqnarray}\label{BHg}
\psi''+(\frac{2}{r}+\frac{g'}{g})\psi'+(\frac{q^2Q^2}{r^2g^2}-\frac{m^2}{g})\psi=0.
\end{eqnarray}

We need to impose boundary conditions to solve the equation (6).
At the star surface, we take the scalar reflecting condition that
the scalar field vanishes. In the region far from the star,
the general solutions behave in the form $\psi\sim A\cdot\frac{1}{r}e^{-m r}+B\cdot\frac{1}{r}e^{m r}$,
where A and B are integral constants.
We set $B=0$ in order to obtain the physical solution
satisfying $\psi(\infty)=0$.
So boundary conditions can be putted as
\begin{eqnarray}\label{InfBH}
&&\psi(r_{s})=0,~~~~~~~~~\psi(\infty)=0.
\end{eqnarray}
With relations (6), (7) and the Hod's method in \cite{Hod-6}, it is easy to check that massive scalar
fields cannot exist outside neutral noncommutative reflecting stars.
In the next section, we turn to study scalar condensations in the background of charged noncommutative reflecting stars.

\section{Scalar field condensations outside noncommutative charged reflecting stars}

\subsection{Upper bounds for radii of noncommutative hairy reflecting stars}

Introducing a new function $\tilde{\psi}=\sqrt{r}\psi$,
we can express the equation (6) as
\begin{eqnarray}\label{BHg}
r^2\tilde{\psi}''+(r+\frac{r^2g'}{g})\tilde{\psi}'+(-\frac{1}{4}-\frac{rg'}{2g}+\frac{q^2Q^2}{g^2}-\frac{m^2r^2}{g})\tilde{\psi}=0.
\end{eqnarray}

The boundary conditions are
\begin{eqnarray}\label{InfBH}
&&\tilde{\psi}(r_{s})=0,~~~~~~~~~\tilde{\psi}(\infty)=0.
\end{eqnarray}
Then the function $\tilde{\psi}$ must possess at least
one extremum point $r=r_{peak}$ above the star radius $r_{s}$.
At this extremum point, the following characteristic relations hold
\begin{eqnarray}\label{InfBH}
\{ \tilde{\psi}'=0~~~~and~~~~\tilde{\psi} \tilde{\psi}''\leqslant0\}~~~~for~~~~r=r_{peak}.
\end{eqnarray}

From (8) and (10), we deduce the inequality
\begin{eqnarray}\label{BHg}
-\frac{1}{4}-\frac{rg'}{2g}+\frac{q^2Q^2}{g^2}-\frac{m^2r^2}{g}\geqslant0~~~for~~~r=r_{peak}.
\end{eqnarray}

It can be transformed into
\begin{eqnarray}\label{BHg}
m^2r^2g\leqslant q^2Q^2-\frac{rgg'}{2}-\frac{1}{4}g^2~~~for~~~r=r_{peak}.
\end{eqnarray}

With the relation $\gamma(s+1,z)=s\gamma(s,z)-x^{s}e^{-z}$,
the metric function $g(r)$ can be expressed as
\begin{equation}\label{BHg}
\begin{split}
g(r)=1-\frac{2M}{\sqrt{\pi}r}[\gamma(\frac{1}{2},\frac{r^2}{4\theta})
-\frac{r}{\sqrt{\theta}}e^{-\frac{r^2}{4\theta}}]
+\frac{Q^2}{\pi r^2}[\gamma^{2}(\frac{1}{2},\frac{r^2}{4\theta})+\frac{r}{\sqrt{2\theta}}
\gamma(\frac{1}{2},\frac{r^2}{4\theta})-\frac{r}{\sqrt{2\theta}}
\gamma(\frac{1}{2},\frac{r^2}{2\theta})
-\frac{r^2}{\sqrt{2}\theta}e^{-\frac{r^2}{4\theta}}].
\end{split}
\end{equation}

Then we have
\begin{equation}\label{BHg}
\begin{split}
g'=\frac{2M}{\sqrt{\pi}r^2}[\gamma(\frac{1}{2},\frac{r^2}{4\theta})-\frac{r}{\sqrt{\theta}}e^{-\frac{r^2}{4\theta}}]
-\frac{2Q^2}{\pi r^3}[\gamma^{2}(\frac{1}{2},\frac{r^2}{4\theta})+\frac{r}{\sqrt{2\theta}}
\gamma(\frac{1}{2},\frac{r^2}{4\theta})-\frac{r}{\sqrt{2\theta}}
\gamma(\frac{1}{2},\frac{r^2}{2\theta})-\frac{r^2}{\sqrt{2}\theta}e^{-\frac{r^2}{4\theta}}]\\
-\frac{2M}{\sqrt{\pi}r}[\frac{2\sqrt{\theta}}{r}e^{-\frac{r^2}{4\theta}}\frac{2r}{4\theta}
-\frac{1}{\sqrt{\theta}}e^{-\frac{r^2}{4\theta}}
+\frac{r^2}{2\theta\sqrt{\theta}}e^{-\frac{r^2}{4\theta}}]
+\frac{Q^2}{\pi r^2}[2\gamma(\frac{1}{2},\frac{r^2}{4\theta})\frac{2\sqrt{\theta}}{r}e^{-\frac{r^2}{4\theta}}\frac{2r}{4\theta}
+\frac{1}{\sqrt{2\theta}}\gamma(\frac{1}{2},\frac{r^2}{4\theta})\\
-\frac{1}{\sqrt{2\theta}}\gamma(\frac{1}{2},\frac{r^2}{2\theta})
+\frac{r}{\sqrt{2\theta}}\frac{2\sqrt{\theta}}{r}e^{-\frac{r^2}{4\theta}}\frac{2r}{4\theta}
-\frac{r}{\sqrt{2\theta}}\frac{\sqrt{2\theta}}{r}e^{-\frac{r^2}{2\theta}}\frac{2r}{2\theta}
-\frac{\sqrt{2}r}{\theta}e^{-\frac{r^2}{4\theta}}
+\frac{r^3}{2\sqrt{2}\theta^2}e^{-\frac{r^2}{4\theta}}].
\end{split}
\end{equation}

And there is also the relation
\begin{eqnarray}\label{BHg}
\gamma(\frac{1}{2},\frac{r^2}{4\theta})-\gamma(\frac{1}{2},\frac{r^2}{2\theta})=
\frac{\partial\gamma(\frac{1}{2},z)}{\partial z}\mid _{z=-\frac{r_{\ast}^2}{4\theta}}(\frac{r^2}{4\theta}-\frac{r^2}{2\theta})
=\frac{2\sqrt{\theta}}{r_{\ast}}e^{-\frac{r_{\ast}^2}{4\theta}}(-\frac{r^2}{4\theta})
\end{eqnarray}
with $r_{\ast}^{2}\in [r^2,2r^2]$.

We mention that $\gamma(\frac{1}{2},\frac{r^2}{4\theta})$ increases as a function
of $r$ with values of $\gamma(\frac{1}{2},\frac{r^2}{4\theta})$ in the range $[0,\sqrt{\pi}]$.
According to (14), (15) and the fact that $e^{-\frac{r^2}{4\theta}}$ decreases very quickly, $g'$
asymptotically behaves as
\begin{equation}\label{BHg}
\begin{split}
g'\thickapprox\frac{2M}{\sqrt{\pi}r^2}\gamma(\frac{1}{2},\frac{r^2}{4\theta})
-\frac{2Q^2}{\pi r^3}\gamma^{2}(\frac{1}{2},\frac{r^2}{4\theta})
\thickapprox\frac{2M}{r^2}
-\frac{2Q^2}{r^3}
\end{split}
\end{equation}
in the large r region.

The dominant terms of (14) satisfy
\begin{equation}\label{BHg}
\begin{split}
\frac{2M}{\sqrt{\pi}r^2}\gamma(\frac{1}{2},\frac{r^2}{4\theta})\geqslant\frac{\frac{3}{2}M}{r^2}~~~for~~~\frac{r^2}{4\theta}\geqslant 1;
\end{split}
\end{equation}
\begin{equation}\label{BHg}
\begin{split}
-\frac{2Q^2}{\pi r^3}\gamma^{2}(\frac{1}{2},\frac{r^2}{4\theta})\geqslant-\frac{2Q^2}{ r^3}.
\end{split}
\end{equation}

Other terms in (14) satisfy
\begin{equation}\label{BHg}
\begin{split}
\frac{2M}{\sqrt{\pi}r^2}[-\frac{r}{\sqrt{\theta}}e^{-\frac{r^2}{4\theta}}]
=-\frac{2M}{r^2}[\frac{2}{\sqrt{\pi}}\frac{r}{2\sqrt{\theta}}e^{-\frac{r^2}{4\theta}}]
\geqslant-\frac{2M}{r^2}\frac{1}{8}~~~for~~~\frac{r^2}{4\theta}\geqslant 6;
\end{split}
\end{equation}
\begin{equation}\label{BHg}
\begin{split}
-\frac{2M}{\sqrt{\pi}r}[\frac{2\sqrt{\theta}}{r}e^{-\frac{r^2}{4\theta}}\frac{2r}{4\theta}
+\frac{r^2}{2\theta\sqrt{\theta}}e^{-\frac{r^2}{4\theta}}]
=-\frac{2M}{r^2}(\frac{2}{\sqrt{\pi}}\frac{r}{2\sqrt{\theta}}e^{-\frac{r^2}{4\theta}}
+\frac{4}{\sqrt{\pi}}\frac{r^3}{8\theta\sqrt{\theta}}e^{-\frac{r^2}{4\theta}})
\geqslant-\frac{2M}{r^2}\frac{1}{8}~~for~~\frac{r^2}{4\theta}\geqslant 6;
\end{split}
\end{equation}
\begin{equation}\label{BHg}
\begin{split}
\frac{Q^2}{\pi r^2}[-\frac{r}{\sqrt{2\theta}}\frac{\sqrt{2\theta}}{r}e^{-\frac{r^2}{2\theta}}\frac{2r}{2\theta}]
=-\frac{Q^2}{r^3}[\frac{2}{\pi}\frac{r^2}{2\theta}e^{-\frac{r^2}{2\theta}}]\geqslant
-\frac{Q^2}{r^3}\frac{1}{3}~~~for~~~\frac{r^2}{4\theta}\geqslant 6;
\end{split}
\end{equation}
\begin{equation}\label{BHg}
\begin{split}
\frac{Q^2}{\pi r^2}[\frac{1}{\sqrt{2\theta}}\gamma(\frac{1}{2},\frac{r^2}{4\theta})
-\frac{1}{\sqrt{2\theta}}\gamma(\frac{1}{2},\frac{r^2}{2\theta})]=\frac{Q^2}{\pi r^2}\frac{1}{\sqrt{2\theta}}
\frac{2\sqrt{\theta}}{r_{\ast}}e^{-\frac{r_{\ast}^2}{4\theta}}(-\frac{r^2}{4\theta})
=-\frac{Q^2}{r^3}\frac{\sqrt{2}}{\pi}\frac{r}{r_{\ast}}\frac{r^2}{4\theta}e^{-\frac{r_{\ast}^2}{4\theta}}\\
\geqslant-\frac{Q^2}{r^3}\frac{\sqrt{2}}{\pi}\frac{r_{\ast}^2}{4\theta}e^{-\frac{r_{\ast}^2}{4\theta}}
\geqslant-\frac{Q^2}{r^3}\frac{1}{3}~~~for~~~\frac{r_{\ast}^2}{4\theta}\geqslant\frac{r^2}{4\theta}\geqslant 6;~~~~~~~~~~~~~~~~~
\end{split}
\end{equation}
\begin{equation}\label{BHg}
\begin{split}
\frac{Q^2}{\pi r^2}[-\frac{\sqrt{2}r}{\theta}e^{-\frac{r^2}{4\theta}}]=
-\frac{Q^2}{r^3}[\frac{4\sqrt{2}}{\pi}\frac{r^2}{4\theta}e^{-\frac{r^2}{4\theta}}]
\geqslant-\frac{Q^2}{r^3}\frac{1}{3}~~~for~~~\frac{r^2}{4\theta}\geqslant 6;~~~~~~~~~~~~~~~~~
\end{split}
\end{equation}
\begin{equation}\label{BHg}
\begin{split}
-\frac{2Q^2}{\pi r^3}[\frac{r}{\sqrt{2\theta}}
\gamma(\frac{1}{2},\frac{r^2}{4\theta})-\frac{r}{\sqrt{2\theta}}
\gamma(\frac{1}{2},\frac{r^2}{2\theta})]\geqslant 0;
\end{split}
\end{equation}
\begin{equation}\label{BHg}
\begin{split}
-\frac{2Q^2}{\pi r^3}[-\frac{r^2}{\sqrt{2}\theta}e^{-\frac{r^2}{4\theta}}]\geqslant 0;
\end{split}
\end{equation}
\begin{equation}\label{BHg}
\begin{split}
-\frac{2M}{\sqrt{\pi}r}[-\frac{1}{\sqrt{\theta}}e^{-\frac{r^2}{4\theta}}]\geqslant 0;
\end{split}
\end{equation}
\begin{equation}\label{BHg}
\begin{split}
\frac{Q^2}{\pi r^2}[2\gamma(\frac{1}{2},\frac{r^2}{4\theta})\frac{2\sqrt{\theta}}{r}e^{-\frac{r^2}{4\theta}}\frac{2r}{4\theta}]\geqslant 0;
\end{split}
\end{equation}
\begin{equation}\label{BHg}
\begin{split}
\frac{Q^2}{\pi r^2}[\frac{r}{\sqrt{2\theta}}\frac{2\sqrt{\theta}}{r}e^{-\frac{r^2}{4\theta}}\frac{2r}{4\theta}]\geqslant 0;
\end{split}
\end{equation}
\begin{equation}\label{BHg}
\begin{split}
\frac{Q^2}{\pi r^2}[\frac{r^3}{2\sqrt{2}\theta^2}e^{-\frac{r^2}{4\theta}}]\geqslant 0.
\end{split}
\end{equation}

According to (14-29), the following relation
\begin{equation}\label{BHg}
\begin{split}
g'>\frac{M}{r^2}-\frac{3Q^2}{r^3}=\frac{M}{r^2}(1-\frac{3Q^2}{Mr})>0
\end{split}
\end{equation}
holds for $r>\frac{3Q^2}{M}$ and $\frac{r^2}{4\theta}\geqslant 6$.

In the case of $\frac{r^2}{4\theta}\geqslant 6$, we have
\begin{equation}\label{BHg}
\begin{split}
\frac{-4\gamma(\frac{3}{2},\frac{r^2}{4\theta})}{\sqrt{\pi}}\geqslant -3,
\end{split}
\end{equation}
\begin{equation}\label{BHg}
\begin{split}
\frac{1}{\pi}[\gamma^{2}(\frac{1}{2},\frac{r^2}{4\theta})
-\frac{r}{\sqrt{2\theta}}\gamma(\frac{1}{2},\frac{r^2}{2\theta})
+\sqrt{\frac{2}{\theta}}r\gamma(\frac{3}{2},\frac{r^2}{4\theta})]\geqslant \frac{1}{2}.
\end{split}
\end{equation}
Then there is $g>1-\frac{3M}{r}+\frac{\frac{1}{2}Q^2}{r^2}$ for $\frac{r^2}{4\theta}\geqslant 6$.
We obtain an lower bound of the metric function as
\begin{equation}\label{BHg}
\begin{split}
g>1-\frac{3M}{r}+\frac{\frac{1}{2}Q^2}{r^2}=\frac{1}{2}+\frac{1}{2}(1-\frac{6M}{r}+\frac{Q^2}{r^2})>\frac{1}{2}
\end{split}
\end{equation}
on conditions $r>3M+\sqrt{9M^2-Q^2}$ and $\frac{r^2}{4\theta}\geqslant 6$.

We assume that the star radius satisfies $\frac{r_{s}^2}{4\theta}\geqslant 6$,
$r_{s}>\frac{3Q^2}{M}$ and $r_{s}>3M+\sqrt{9M^2-Q^2}$, otherwise we will
arrive at an upper bound
\begin{equation}
mr_{s} \leqslant max \left\{2m\sqrt{6\theta},\frac{3mQ^2}{M},3mM+m\sqrt{9M^2-Q^2} \right\}.
\end{equation}

According to (12), (30), (33) and $r_{peak}\geqslant r_{s}$, there is
\begin{eqnarray}\label{BHg}
\frac{1}{2}m^2r_{s}^2 \leqslant \frac{1}{2}m^2r^2 \leqslant m^2r^2g\leqslant q^2Q^2-\frac{rgg'}{2}-\frac{1}{4}g^2\leqslant q^2Q^2-\frac{1}{4}g^2\leqslant q^2Q^2-\frac{1}{16}~~~for~~~r=r_{peak}.
\end{eqnarray}

Then we have
\begin{eqnarray}\label{BHg}
m^2r_{s}^2 \leqslant 2q^2Q^2-\frac{1}{8}.
\end{eqnarray}

In all, our analysis shows that the hairy star radius is below the bounds (34) or (36).
With dimensionless quantities according to the symmetry (38), we obtain upper bounds for hairy star radii as
\begin{equation}
mr_{s} \leqslant max \left\{2m\sqrt{6\theta},\frac{3mQ^2}{M},3mM+m\sqrt{9M^2-Q^2},\sqrt{2q^2Q^2-\frac{1}{8}} \right\}.
\end{equation}
Above this bound, the static scalar field cannot condense.
Below the bound, we will numerically obtain hairy reflecting star solutions
in the next part.

We mention that there are also upper bounds on the size of hairy black holes.
According to the no short hair conjecture, the hairy black hole
has an upper bound for the horizon $r_{H}<\frac{2}{3}(\eta)^{-1}$,
where $\eta$ is the field mass \cite{ub1}.
In fact, the numerical results in \cite{ub2} also support the
idea that big black holes tend to have no massive hair.
Similar to black hole theories, we find that big reflecting star cannot support
the existence massive scalar hair.
So it is natural that there is a maximal
radius for the hairy star. In the following,
we will numerically search for the maximal radius.

\subsection{Scalar field configurations supported by noncommutative charged reflecting stars}

We firstly show the metric solution $g(r)$ with different values of the noncommutative
parameter $\theta$ in Fig. 1.
It can be seen from the panels that the noncommutative spacetime is regular at $r=0$
and the metric behaves like Schwarzschild geometry at $r\rightarrow\infty$,
in accordance with results in \cite{PN5}.
Our results also imply that in cases of very small $\theta$,
the metric almost coincides with the Schwarzschild
geometry even at the points $r\thickapprox0$,
but the metric always keeps regular at the center with $g(0)=1$.
We point out that the regular star radius should be imposed
above the outmost would-be horizon of the metric.

\begin{figure}[h]
\includegraphics[width=180pt]{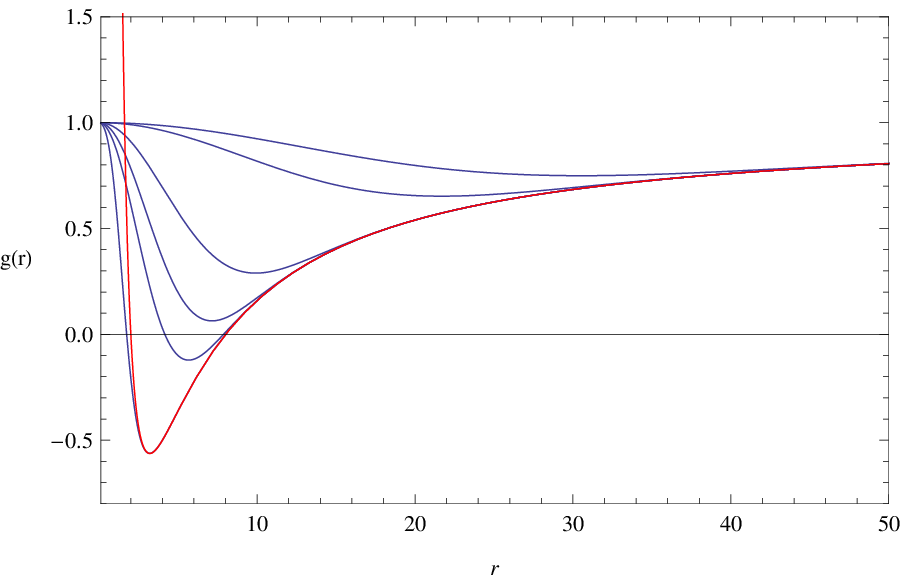}\
\includegraphics[width=175pt]{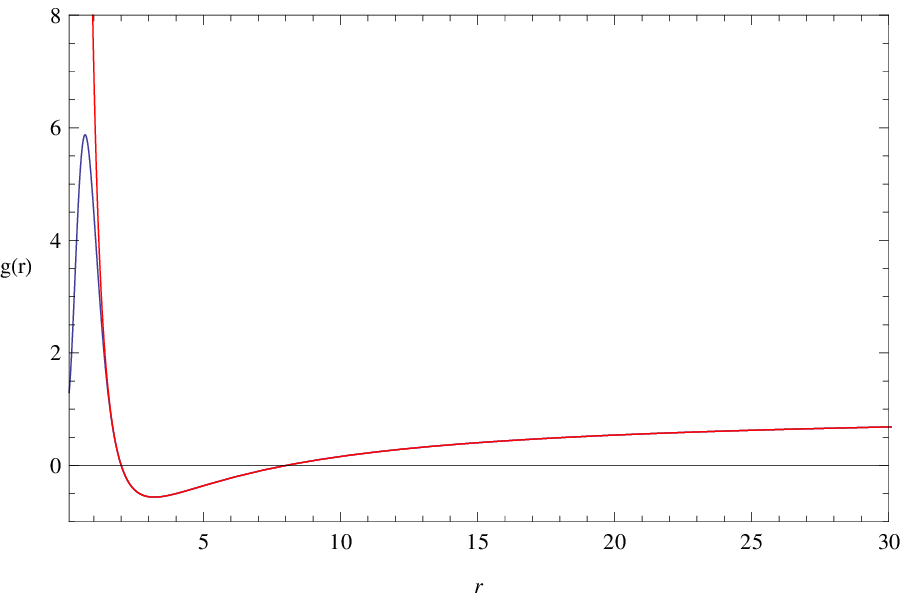}\
\caption{\label{EEntropySoliton} (Color online) We plot $g(r)$ as a function of
the coordinate $r$ with $q=2$, $Q=4$ and $M=5$.
In the left panel, the blue curves from top to bottom correspond to various $\theta$
as $\theta=100$, $\theta=50$, $\theta=10$, $\theta=5$, $\theta=3$ and $\theta=0.5$.
The blue curve in the right panel is the case of $\theta=0.1$.
And the red lines represent Schwarzschild geometries with $\theta=0$.}
\end{figure}

In the following, we numerically obtain regular configurations composed of
scalar fields and noncommutative charged reflecting stars.
We take $m=1$ according to the symmetry
\begin{eqnarray}\label{BHg}
r\rightarrow k r,~~~~ m\rightarrow m/k,~~~~ M\rightarrow k M,~~~~ Q\rightarrow k Q,~~~~ q\rightarrow q/k,~~~~ \theta\rightarrow k^2 \theta.
\end{eqnarray}
Around the star surface, the scalar field behaves in the form $\psi=\psi_{0}(r-r_{s})+\cdots$.
With the symmetry $\psi\rightarrow k \psi$ of equation (6),
we can set $\psi_{0}=1$ without loss of generality.
Then we numerically search for the proper $r_{s}$,
where the corresponding scalar field satisfies the
boundary condition (7) at the infinity.

In Fig. 2, we plot scalar fields in the background of compact stars
with $q=2$, $Q=4$, $M=5$ and $\theta=50$.
With the star radius fixed at $r_{s}= 9.6719$,
the scalar field approaches zero at the infinity.
If we impose the star radius a little larger or a little
smaller than $r_{s}$, the solutions asymptotically behave
in the form $\psi\varpropto A\cdot\frac{1}{r}e^{-m r}+B\cdot\frac{1}{r}e^{m r}$ with nonzero B,
which contradicts the infinity boundary condition (7).
Our detailed calculations show that the hairy star
radius is discrete.

\begin{figure}[h]
\includegraphics[width=210pt]{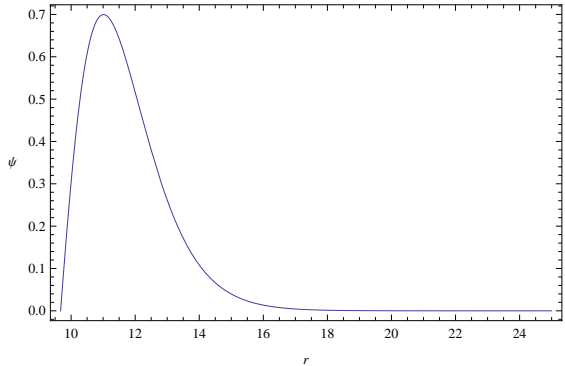}\
\caption{\label{EEntropySoliton} (Color online) We show $\psi$ as a function of
the coordinate $r$ with $q=2$, $Q=4$, $M=5$, $\theta=50$ and the star radius at $r_{s}=9.6719$.}
\end{figure}

Integrating the equation from $r_{s}=9.6719$ to smaller radial
coordinates, we find discrete points,
which can be fixed as the hairy star radii.
In Fig. 3, the discrete points are around $m r_{s}\thickapprox 9.672,~~8.028,~~6.867, \cdots$
below the upper bound $mr_{s}\leqslant
max \left\{2\sqrt{300},\frac{48}{5},15+\sqrt{9\cdot25-16},\sqrt{128-\frac{1}{8}} \right\}
=2\sqrt{300}\thickapprox34.641$ according to (37).
Similar to cases in commutative spacetimes \cite{Hod-8,Hod-9,Yan Peng-1,Hod-10,Yan Peng-2,Yan Peng-3,Yan Peng-4},
we find many discrete hairy star radii.

\begin{figure}[h]
\includegraphics[width=210pt]{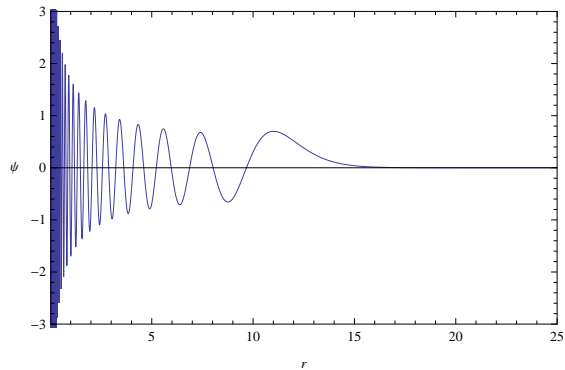}\
\caption{\label{EEntropySoliton} (Color online) We show various discrete radii
with $q=2$, $Q=4$, $M=5$, $\theta=50$ and the largest hairy star radius at $r_{s}=9.6719$.}
\end{figure}

In the case of $q=2$, $Q=4$, $M=5$ and $\theta=50$,
we numerically find that $m r_{s}=9.6719$ is the largest hairy star radius
below the upper bound $m r_{s}\leqslant 2\sqrt{300}\thickapprox34.641$ of (37).
For every given set of parameters, we can obtain the largest hairy star radius labeled as $m R_{s}$.
We also study effects of the noncommutative parameter on the largest hairy star radius.
With dimensionless quantities according to the symmetry (38),
we plot $m R_{s}$ as a function of $m^2\theta$ with $q=2$, $Q=4$ and $M=5$ in Fig. 4.
It can be seen from the picture that larger $m^2\theta$ corresponds to smaller $m R_{s}$.

\begin{figure}[h]
\includegraphics[width=210pt]{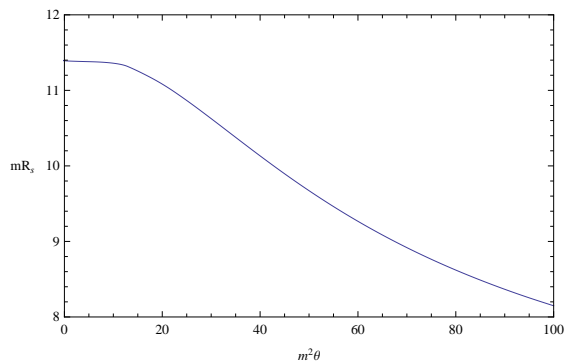}\
\caption{\label{EEntropySoliton} (Color online) We show the largest hairy star radius
with $q=2$, $Q=4$, $M=5$ and various noncommutative parameters.}
\end{figure}

The noncommutative physics is expected to be detected for a small size star
$r_{s}\backsimeq \sqrt{\theta}$.
However, the phenomenological impact of these results
may be not visible since the presently accessible energy
is $\sqrt{\theta}<10^{-16}$ cm \cite{PN5}.
For large distance, the solution behaves very similar to the Schwarzschild metric.
Recently, holographic superconductor models were constructed in the background of
noncommutative AdS black holes \cite{ph1,ph2},
which provided a novel way to investigate
the role of noncommutative geometry through the AdS/CFT duality.

\section{Conclusions}

We studied static scalar field condensations outside noncommutative charged compact reflecting stars.
We found that the scalar field cannot condense outside
neutral noncommutative reflecting stars.
And in the  background of charged noncommutative reflecting compact stars, we provided upper bounds for the star radius
as $mr_{s} \leqslant max \left\{2m\sqrt{6\theta},\frac{3mQ^2}{M},3mM+m\sqrt{9M^2-Q^2},\sqrt{2q^2Q^2-\frac{1}{8}} \right\}$,
where $m$ is the scalar field mass, q is the charge coupling parameter, M is the ADM mass, Q is
the star charge and $\theta$ is the noncommutative parameter.
Above the bound, the scalar field cannot condense
and below the bound, we obtained numerical solutions of scalar hairy reflecting stars.
With detailed calculations, we found that the scalar hairy reflecting star radius is discrete.
We also examined effects of the noncommutative parameter on the largest radius
of the scalar hairy reflecting star.

\begin{acknowledgments}

We would like to thank the anonymous referee for the constructive suggestions to improve the manuscript.
This work was supported by the Shandong Provincial Natural Science Foundation of China under Grant
Nos. ZR2018QA008 and ZR2017BA006. This work was also supported by the National Natural Science
Foundation of China under Grant Nos. 11775076,11703015 and
11690034; Hunan Provincial Natural Science Foundation of China under Grant No. 2016JJ1012.

\end{acknowledgments}

\end{document}